\documentclass[%
reprint,
amsmath,amssymb,
aps, physrev,
nobalancelastpage,
longbibliography,
]{revtex4-2}
 
\usepackage{graphicx}
\usepackage{dcolumn}
\usepackage{bm}
\usepackage{tikz}
\usepackage[labelformat=simple]{subcaption}
\usetikzlibrary{calc}
\usepackage{float}
\usepackage{comment}
\usepackage{hyperref}
\usepackage{cleveref}
 
\newcommand{\customcaption}[5]{{%
  \begin{tikzpicture}
      \node[anchor=north west,inner sep=0] (image) at (0.1,-0.2) {\includegraphics[#1]{#2}};
  \begin{scope}[x={(image.north east)},y={(image.south west)}]
    \draw[#5] (0,0) node [font=\footnotesize, anchor=north west] {{\textbf{#3}}};
  \end{scope}
  \end{tikzpicture}%
  \phantomsubcaption\label{#4}%
}}
 
\usepackage{listings}
\usepackage{xcolor}
 
\definecolor{codegreen}{rgb}{0,0.6,0}
\definecolor{codegray}{rgb}{0.5,0.5,0.5}
\definecolor{codepurple}{rgb}{0.58,0,0.82}
\definecolor{bgcolor}{rgb}{0.96,0.95,0.92}
\definecolor{probcolor}{rgb}{0,0,0}
\definecolor{epflblue}{HTML}{007480}

\hypersetup{
colorlinks=true,
linkcolor=blue,
filecolor=magenta,
urlcolor=blue,
}
 
\begin{document}
 
\title{\textbf{Training of scattering neural networks based on coupled electromagnetic dipoles}} 
\author{Tim Tuuva}

\author{Romain Fleury}%
\affiliation{Laboratory of Wave Engineering, EPFL, 1015 Lausanne, Switzerland}

\date{\today}
 
\begin{abstract}
Physical neural networks are dynamical systems with adjustable weights that leverage the complexity of nature in order to perform efficient analog computation, aiming to demonstrate alternatives to power-hungry digital AI architectures.
Finding the optimal platform for such computation\textemdash one that pairs dedicated hardware with a tailored training algorithm\textemdash remains an open challenge.
Here, we explore wave-based neural networks built from $N$ resonant scatterers coupled by electromagnetic waves, such as loaded dipole antennas.
The inputs and weights are encoded in the detuning of antenna resonances, inducing a form of structural nonlinearity that is enhanced by the non-locality of the network response, enabled by multiple scattering.
To perform gradient descent, we adapt the Equilibrium Propagation (EP) algorithm and particularize it to this specific driven-dissipative system, showing that model-free gradient extraction is possible by performing two measurements per tunable antenna, using reciprocity and the diagonal dependence of the interaction operator.
Since the network is read out in intensity, the cost is not a holomorphic function of the complex dipole moments, and existing holomorphic formulations of EP do not apply.
We derive a modified EP learning rule showing that the nudged phase captures the holomorphic part of the gradient, and that taking its real part recovers the exact gradient.
We numerically implement a compact model with fewer than $7000$ parameters that achieves over $94\%$ test accuracy on MNIST digit classification, and over $81\%$ on Fashion-MNIST classification.
We quantify the robustness of the training method to noise and spatial disorder.
\end{abstract}

\maketitle
  
\section{Introduction}
 
The current growth of artificial intelligence (AI) has led to an exponential increase in power consumption from dedicated data centers~\cite{Sajadieh2026}.
Realizing similar computation at a lower environmental cost has become increasingly important.
This has revived the field of  analog computing as a promising solution for information processing in AI~\cite{Kalinin2025}, with the goal of leveraging intrinsic parallelism, low latency and reduced power consumption compared to digital systems.
Various physical systems have been explored, such as mechanical~\cite{Wright2022a, Li2024}, electronic and CMOS~\cite{Medina-Santiago2021, Dillavou2022, Zhu2023}, radio-frequency~\cite{Sol2024}, and photonic circuits or diffractive optical neural networks~\cite{Lin2018b, Guo2021, Yan2022, Stroev2023, Xu2024}.
The latter are particularly interesting for their many degrees of freedom and their scalability~\cite{Zhou2021, Valensise2022, Chen2023a, McMahon2023}.
Wave-based systems exploit interference, resonances, and multiple scattering to perform high-dimensional transformations of input signals, while learning consists of adjusting physically accessible parameters to achieve a desired input--output mapping.
 
The major challenge for physical neural networks is the development of training strategies that are compatible with the underlying dynamics.
Analog computing is subject to noise and other experimental errors, such as manufacturing defects, component mismatches, limited sensor precision, and environmental drift, which need to be accounted for in the training scheme.
Gradients computed on a digital model and applied to the hardware inherit the discrepancy between the model and its physical implementation, which limits \textit{in silico} training with backpropagation.
Alternatively, learning paradigms inspired by physics and neuroscience have been proposed to address these limitations.
Among these, Equilibrium Propagation (EP)~\cite{Scellier2017a} provides a learning framework in which parameter updates are obtained by comparing two nearby equilibria of a potentially nonlinear dynamical system: a free phase and a nudged phase, corresponding to the unperturbed state and a weakly perturbed one with slightly lower cost.
EP has attracted attention due to its reliance on local information, its compatibility with a broad range of loss functions, and its potential compatibility with nonlinear analog hardware~\cite{Kendall2020, Rageau2025a}.
As a digital training method, it has been benchmarked at large scales, showing accuracies comparable to backpropagation~\cite{Laborieux2021,Kerjan2026}.
Yet, for physical systems, the implications of EP, such as how to nudge the system or which quantities must be measured to extract the gradient, are largely system-dependent.
 
Several works have already exploited the physics of the hardware itself to obtain gradients.
In electromagnetism, the adjoint method enables efficient gradient computation for inverse design governed by Maxwell's equations~\cite{Molesky2018}, and adjoint-based techniques have been applied to the training of physical neural networks~\cite{So2020, Ma2021}.
Hughes \textit{et al.}~\cite{Hughes2018} showed that, in reciprocal photonic circuits, the device can generate its own adjoint field, so that gradients follow from measurements of the forward and adjoint fields;
Pai \textit{et al.}~\cite{Pai2023} later demonstrated this experimentally, and B\"osch \textit{et al.}~\cite{Bosch2026} recently unified such physical-adjoint schemes with EP.
In parallel, EP has been extended beyond conservative, real-valued systems.
Scellier \textit{et al.}~\cite{scellier2022agnostic} showed that EP gradients can be extracted agnostically, without a model of the physical system, and Laborieux and Zenke~\cite{Laborieux2022} introduced holomorphic EP, in which a holomorphic energy and a holomorphic cost yield exact gradients from finite oscillating nudges.
Finally, Wanjura and Marquardt~\cite{Wanjura2024a}  showed that encoding the input in the parameters of a linear scattering system, rather than in the incident field, produces a nonlinear input--output map.
 
The present work builds on these ingredients and addresses a gap between them.
Wave-based networks are usually read out in intensity, so the cost is a real-valued function of the complex state and is not holomorphic; the holomorphic-EP identity then no longer applies.
Our first contribution is to show that, for a holomorphic energy and any real-valued cost, the nudged phase captures only the holomorphic part of the gradient, while the missing antiholomorphic part is its complex conjugate, so that the exact gradient is recovered by taking the real part of the EP estimator.
For linear reciprocal systems, this result reduces to the \textit{in situ} adjoint gradient of Hughes \textit{et al.}~\cite{Hughes2018}, a connection we make explicit.
Our second contribution is to identify networks of coupled resonant dipoles as a platform where this scheme is particularly simple: the trainable detunings $\theta_i$ enter the interaction operator $M$ only through its diagonal, $\partial M/\partial\theta_i = |i\rangle\langle i|$, so each update requires only the currents measured on the $i$-th element in the free and adjoint phases, a result that extends formally to antennas with holomorphic nonlinear responses (Appendix~\ref{app:nonlinear}).
 
We apply this framework to a network of $N$ coupled electromagnetic dipoles, modeled as driven and passive line currents with resonant polarizabilities interacting through multiple scattering, a configuration that captures the essential physics of reconfigurable metasurfaces and antenna arrays~\cite{Hammami2026}.
Inputs and weights are both encoded in the resonance detunings, i.e., in the real part of the inverse polarizability.
Although the field equations are linear, the scattered field depends nonlinearly on these parameters through the inverse of the coupled-dipole operator (Sec.~\ref{sec:Theory}), which allows the network to perform nonlinear classification tasks.
Through numerical examples with MNIST and Fashion-MNIST, we illustrate learning and classification tasks in such coupled-dipole networks under realistic passivity constraints.
We achieve test accuracies over $94\%$ on MNIST and over $81\%$ on Fashion-MNIST with fewer than 7000 parameters, comparable to recent wave-based physical neural networks of similar scale~\cite{Thakkar2026}.
Finally, we characterize the effect of additive measurement noise and random positional disorder, which are crucial for a physical implementation (Sec.~\ref{sec:result-noise}).

\section{Theory}\label{sec:Theory}
\subsection{Coupled-dipole model in two-dimensional electromagnetism}\label{sec:cpdl}
We consider a two-dimensional electromagnetic system invariant along the z direction, corresponding to a transverse-magnetic (TM) polarization with nonzero out-of-plane electric field component $E_z$ (convention used in 2D photonic crystals~\cite{joannopoulos2008photonic}).
The system consists of $N$ infinitely long line currents oriented along z, located at positions $\{\mathbf{r}_i\}_{i=1}^N$ in the $(x,y)$ plane, representing the currents flowing through N polarizable elements, hereafter called antennas.
Each antenna is modeled as an electric line dipole characterized by a scalar polarizability $\alpha_i$, relating its induced dipole moment $p_i$ to the local electric field.
At a fixed angular frequency $\omega$ (with the time convention $e^{j\omega t}$), the induced dipole moments satisfy the coupled-dipole equations \cite{DeVoe1964}
 
\begin{equation}\label{eq:couple-dipole}
    p_i = \alpha_i\left(E_i + \sum_{j\neq i}G(\mathbf{r}_i,\mathbf{r}_j)p_j\right),
\end{equation}
where $E_i$ is the externally applied field driving the $i$-th antenna (which can be local, due to a source, or due to an incident wave field) and $G(\mathbf{r}_i, \mathbf{r}_j)$ describes the field at $\mathbf{r}_i$ created by a dipole $p_j$ at $\mathbf{r}_j$.
It is the two-dimensional scalar Green's function of the Helmholtz equation:
 
\begin{equation}
    G(\mathbf{r}_i, \mathbf{r}_j) = -j\frac{k^2}{4\epsilon} H_0^{(2)}(k|\mathbf{r}_i-\mathbf{r}_j|),
\end{equation}
with $k=\sqrt{\epsilon}\ \omega/c$, where $\epsilon$ is the permittivity of the background medium and $c$ is the vacuum speed of light.
 
Eq. (\ref{eq:couple-dipole}) can be written in a matrix form as

\begin{equation}\label{eq:compact-cpldip}
   \left(\alpha^{-1} - G\right)p = Mp= E,
\end{equation}
where $\alpha=\text{diag}(\alpha_1,\cdots,\alpha_N)$ and G is the interaction matrix with elements $G_{ij}=G(\mathbf{r_i}, \mathbf{r_j})$ and a vanishing diagonal.
The system is also described by the stationary functional:
 
\begin{equation} \label{eq:potential}
    \mathcal{E}(\theta, p) = \frac{1}{2}p^\top Mp - p^\top E,
\end{equation}
which is holomorphic in $p$.
We denote by $p^{(0)}$ the stationary point of $\mathcal{E}$, which corresponds to the state associated with the ‘‘prediction’’ $\hat{y}$ of our physical neural network model.
The state Eq.~(\ref{eq:compact-cpldip}) is recovered from the stationarity condition:
\begin{equation}\label{eq:state_lin}
    \frac{\partial \mathcal{E}}{\partial p} = 0 \Rightarrow Mp=E.
\end{equation}
Since $G(\mathbf{r_i}, \mathbf{r_j}) = G(\mathbf{r_j}, \mathbf{r_i})$, by reciprocity of the Helmholtz Green's function, the interaction matrix $G$ and hence $M$ are symmetric: $M=M^\top$.
This satisfies the Schwarz/Poincaré lemma condition: a holomorphic vector field $f(p)=Mp-E$  is the gradient of some scalar if and only if its Jacobian is symmetric everywhere.
 
\textit{Remark:}
The construction above is not restricted to linear antennas.
If each antenna has an additional amplitude-dependent response that is a holomorphic function of its own dipole moment, the state equation still derives from a holomorphic functional with a symmetric Jacobian, and the training scheme of Sec.~\ref{sec:EP} carries over with $M$ replaced by this Jacobian.
We give this formal extension in Appendix~\ref{app:nonlinear}.
In the rest of the paper, and in all simulations, we restrict ourselves to linear scattering.
 
Moreover, wave energy conservation imposes a constraint on the imaginary part of the inverse complex polarizability, while the real part encodes the resonance.
Indeed, under a Lorentz model assumption with radiation loss $\Gamma_{rad}$, the real part of the inverse polarizability is proportional to the frequency detuning between the operating frequency $\omega$ and the dipolar resonance frequency $\omega_i$:
 
\begin{equation}
    \operatorname{Re}(\alpha_i^{-1}) = \frac{k^2}{2\epsilon}\frac{(\omega_i - \omega)}{\Gamma_{rad}}.
\end{equation}
 
In the two-dimensional case, the optical theorem requires that each dipole satisfies \cite{Sipe1974}
 
\begin{equation}\label{eq:passivity}
    \operatorname{Im}(\alpha_i^{-1}) = \frac{k^2}{4\epsilon} \left(\frac{\Gamma_{rad} + \Gamma_{nr}}{\Gamma_{rad}}\right) \geq \frac{k^2}{4\epsilon},
\end{equation}
where equality corresponds to a lossless but radiating antenna, and strict inequality accounts for additional absorption by the scatterer.
This radiative contribution is universal and ensures passivity of the coupled 2D system.
As a consequence of Eq.~(\ref{eq:passivity}), the operator $M$ is non-Hermitian, and when it is invertible, the steady-state solution of Eq.~(\ref{eq:compact-cpldip}) is unique and reads
\begin{equation}\label{eq:cpldip-sol}
    p^{(0)} =M^{-1}E,
\end{equation}
which defines the time-harmonic steady state of the system.
All electromagnetic observables of interest, including near fields, far-field radiation, and currents on selected receiving antennas, can then be expressed as linear functionals of $p^{(0)}$.
If we assume that $E^{i}$ is due to local sources exciting the antennas, and not an incident external field, the electric field at any location $\mathbf{r}$ can be obtained from:
\begin{equation}
    E(\mathbf{r}) = \sum_{i=1}^N G(\mathbf{r}, \mathbf{r}_i)p_i,
\end{equation}
 
where we dropped the superscript $^{(0)}$ for convenience of notation. In the following, the steady-state dipole moments $p$ are interpreted as the internal state of a wave-based computational system, while learning consists of adjusting the weights, represented by the detunings $\operatorname{Re}(\alpha_i^{-1})$.
We express $\alpha^{-1}$ and $G$ in units of $\frac{k^2}{4\epsilon}$, so that $\operatorname{Im}(\alpha_i^{-1}) = 1 + \frac{\Gamma_{nr}}{\Gamma_{rad}}$ and $\operatorname{Re}(\alpha_i^{-1}) = 2\frac{(\omega_i - \omega)}{\Gamma_{rad}}$.
In all simulations, we use $\frac{\Gamma_{nr}}{\Gamma_{rad}} = 0.01$.
 
\subsection{Model-free algorithm for training} \label{sec:EP}
 
We now derive a model-free training scheme with respect to the internal scattering operator $M$.
This method can be pictured either as a special instance of the adjoint method \cite{Giles2000a} in which the same system can be used to determine the adjoint field (thanks to reciprocity), or as a version of Equilibrium Propagation (EP) \cite{Scellier2017a} particularized to a linear system under a driven-dissipative equilibrium (a steady state).
 
\begin{figure}[t]
    \centering
    \includegraphics[width=0.48\textwidth]{"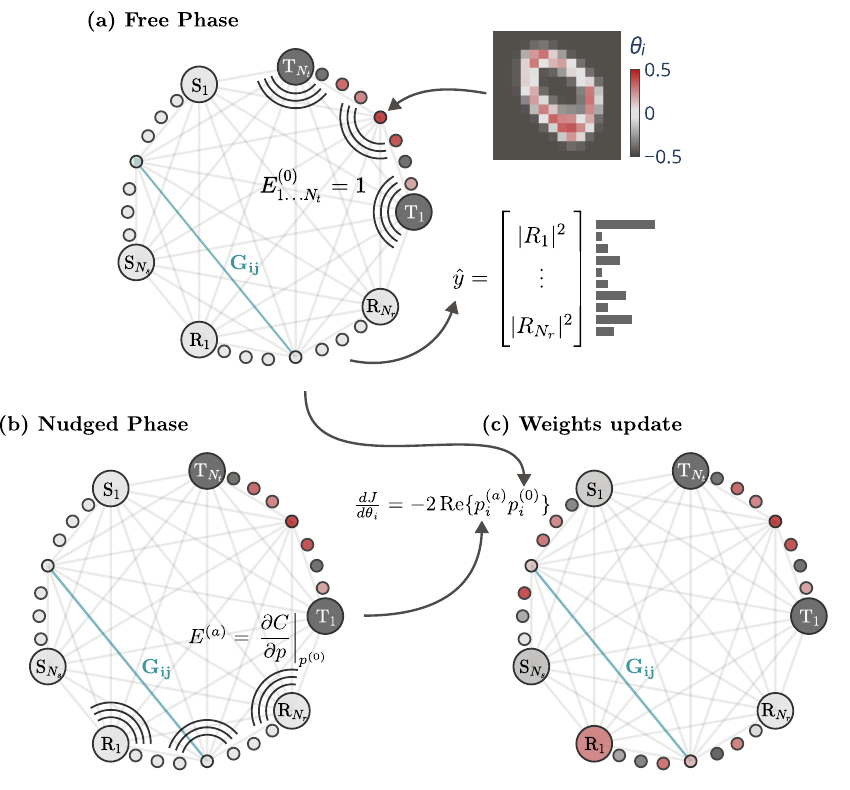"}
    \caption{
        \small{Complete graph representation of our physical neural network, consisting of three types of antennas (Transmitters (\textbf{T}), Scatterers (\textbf{S}), and Receivers (\textbf{R})). The input is encoded in the frequency detuning of the emitting antennas.
        \textbf{(a)} During the free phase, only the emitting antennas are driven, with a harmonic excitation of constant phase and amplitude $E^{(0)}_{1\dots N_t} = 1$. Once the system reaches steady state, we measure the intensity on every receiver, where the largest intensity corresponds to the prediction $\hat{y}$.
        \textbf{(b)} During the nudged phase, the $N_r$ receiver dipoles are driven harmonically with a different excitation $E^{(a)}$, called the adjoint excitation, while the input remains encoded in the transmitters' detunings. Once steady state is reached, we measure the dipole moments $p^{(a)}$ in every element.
        \textbf{(c)} The gradient of the loss is computed from the \textit{free} and \textit{adjoint} dipole moments, namely $p^{(0)}$ and $p^{(a)}$, such that $\dfrac{d J}{d \theta_i} = -2\operatorname{Re}\{p^{(0)}_ip^{(a)}_i\}$. The updated parameters are then incorporated in the detuning of every scattering and receiving dipole.
        }
    }
    \label{fig:cpldip_ex}
\end{figure}

Let us define an array of $N = N_{t} + N_{s} + N_{r}$ coupled dipoles as shown in Fig.~\ref{fig:cpldip_ex}, with  $N_{t}$, $N_{s}$, and $N_{r}$ the numbers of transmitters, scatterers, and receivers, respectively.
We define the trainable parameters $\mathbf{\theta}$ to be the normalized dipole detunings
\begin{equation}
    \theta_i \equiv \operatorname{Re}(\alpha_i^{-1}),
\end{equation}
and consider a dataset composed of input data $x$ and labels $y$, from which our neural network predicts $\hat{y}$.
The input data $x$ are normalized to $[-0.5, 0.5]$, spanning half of a resonance linewidth.
They are then encoded in the frequency detuning of the $N_t$ emitting dipoles as $\theta_i = x_i$ for $i \in \{1,\ldots, N_t\}$, which is a form of structural nonlinearity encoding \cite{Momeni2023, Sol2023, Xia2024, Wanjura2024a, Yildirim2024a}.
 
Notice that the operator $M$ contains all the parameters $\theta_i$ exclusively as linear diagonal terms, so that:
\begin{equation}\label{eq:M_theta}
    \dfrac{\partial M}{\partial \theta_i} = |i\rangle\langle i|.
\end{equation}
This is where the specific coupled-dipole model turns out to be favorable for physical implementation, as other systems with a linear state law may not exhibit such a diagonal embedding of the weights.
For example, in networks whose tunable elements are the couplings themselves (e.g., series elements between nodes, or phase shifters in interferometric meshes), the parameters enter off-diagonal or multiplicatively.
In an experimental implementation, the detuning may be achieved with a reactive load (a varactor) that adds an imaginary impedance, which changes only $\operatorname{Re}(\alpha_i^{-1})$ and leaves the radiative imaginary part untouched.
This property is what makes this model-free training scheme practically relevant, and it can be found in other physical neural network systems with similar mathematical structures
 \cite{Bosch2026}.
 
We perform two forward passes to train the network, namely the \emph{free phase}, with dipole moments $p^{(0)}$ and electric field $E^{(0)}$, and the \emph{nudged phase}, with dipole moments $p^{(\beta)}$ and electric field $E^{(\beta)}$.
The only difference between the free and nudged phases is that, while the free phase records the prediction of the network at the current training stage, the nudged phase includes additional excitations such that the gradient can be estimated exactly by measuring $p^{(0)}$ and $p^{(\beta)}$, and, crucially, without knowing or measuring $M$.
The parameter $\beta$ is the nudging strength.
The EP gradient below is defined in the limit $\beta\to0$.
We show that, for the linear system considered here, the nudge can be implemented such that the result holds exactly for any finite $\beta$, which then drops out.
In the nonlinear extension of Appendix~\ref{app:nonlinear}, $\beta$ must instead be kept small, so that $p$ responds linearly to the nudge.
The inference pass is performed during the free phase, the emitting antennas are driven with a harmonic excitation of constant phase and amplitude, the prediction $\hat{y}$ corresponds to the receiver with the largest intensity when the system has reached steady state.
The receivers' intensities are retrieved from all the dipoles with a selection matrix $W\in \{0,1\}^{N_{r}\times N}$:
\begin{equation}
    \hat{y} = |Wp^{(0)}|^2.
\end{equation}
 
Our goal is to minimize a given loss $J(y, \hat{y})$ (or objective function) without any knowledge of our physical system, defined by the operator $M(\theta)$.
We define the cost function $C(\theta, y, p)$, related to the loss function by:
\begin{equation}\label{eq:loss_cost}
    J(y, \hat{y}) := C(\theta, y, p^{(0)}).
\end{equation}
Thus, the loss function $J$ implicitly depends on $\theta$ through the prediction $\hat{y}$.
The cost $C(\theta,y,p)$ is real-valued and depends on $p$ only through the receiver intensities $|Wp|^2$, with no explicit dependence on $\theta$.
It therefore depends on both $p$ and $p^*$, so $C$ is not holomorphic.
 
Recall that vanilla gradient descent with a learning rate $\eta>0$ for the parameter $\theta_i$ at step $k$ reads
\begin{equation}
    \theta_i^k = \theta_i^{k-1} - \eta\frac{d J}{d \theta_i}.
\end{equation}
 
To compute this gradient and minimize the loss $J(y, \hat{y})$, we exploit the EP framework by extending the notion of a fixed point from our initial Eq.~(\ref{eq:potential}) to any value of $\beta$ with the total function:
\begin{equation}
    F(\theta, y, \beta, p) = \mathcal{E}(\theta, p) + \beta C(\theta, y, p)
\end{equation}
$F$ inherits the $p^*$-dependence of $C$.
It is therefore neither holomorphic nor real-valued, so it has no minimum.
Moreover, the nudged state cannot be defined as a stationary point of $F$ in the full $(p,p^*)$ sense, because $\partial F/\partial p^*=\beta\,\partial C/\partial p^*\neq 0$ for $\beta\neq 0$.
We instead define the nudged state $p^{(\beta)}$ by the Wirtinger condition $\partial F/\partial p=0$, that is, by the driven equation:
\begin{align}
    &\left.\frac{\partial F}{\partial p}\right|_{p^{(\beta)}} = 0 \\
    &Mp^{(\beta)} - E^{(0)} + \beta \left. \frac{\partial C}{\partial p}\right|_{p^{(\beta)}}  = 0 \\
    &p^{(\beta)} = M^{-1}\underbrace{\left(E^{(0)} - \beta \left. \frac{\partial C}{\partial p}\right|_{p^{(\beta)}}\right)}_{=E^{(\beta)}},
\end{align}
hence:
\begin{equation}\label{eq:nudging}
    \delta p = p^{(\beta)} - p^{(0)} = -\beta M^{-1}\left.\frac{\partial C}{\partial p}\right|_{p^{(0)}} + \mathcal{O}(\beta^2),
\end{equation}
where we used $\partial C/\partial p|_{p^{(\beta)}} = \partial C/\partial p|_{p^{(0)}} + \mathcal{O}(\beta^2)$.
Only the first-order term enters the gradient formula below.
In practice, we apply the nudge with the excitation frozen at the free state, $E^{(\beta)} = E^{(0)} - \beta\,\partial C/\partial p|_{p^{(0)}}$, which only requires quantities measured in the free phase.
Since the system is linear, this frozen nudge yields exactly $\delta p = -\beta M^{-1}\partial C/\partial p|_{p^{(0)}}$ for any $\beta$, with no higher-order terms.
The gradient defined by Scellier and Bengio in EP \cite{Scellier2017a}, and modified by Laborieux and Zenke for holomorphic networks \cite{Laborieux2022}, has to be adapted in our framework for non-holomorphic readout (detailed derivation in Appendix~\ref{app:EP_nonholo}):
\begin{equation}\label{eq:ep_nonholo}
    \frac{d J}{d \theta} = \lim_{\beta \to 0}\frac{2}{\beta}\operatorname{Re}\left\{\frac{\partial F}{\partial \theta}(\theta,y,\beta,p^{(\beta)}) - \frac{\partial F}{\partial \theta}(\theta,y,0,p^{(0)})\right\}.
\end{equation}
This expression simplifies (see Appendix~\ref{app:EP_derivation}) to:
\begin{equation} \label{eq:update_rule}
    \frac{d J}{d \theta_i} = -2\operatorname{Re}\left\{p^{(0)\top}|i\rangle\langle i|M^{-1}\frac{\partial C}{\partial p}\bigg|_{p^{(0)}}\right\}.
\end{equation}
With the frozen nudge, $\delta p/\beta$ is independent of $\beta$, so Eq.~(\ref{eq:update_rule}), which depends on the nudged phase only through $\delta p/\beta$, is exact for any finite $\beta$ and the weight update does not depend on the nudging strength.
Evaluated directly at finite $\beta$, Eq.~(\ref{eq:ep_nonholo}) would contain an additional $\mathcal{O}(\beta^2)$ term inside the braces, i.e., an $\mathcal{O}(\beta)$ bias on the gradient; this bias vanishes with a symmetric nudge $\pm\beta$, as in Appendix~\ref{app:nonlinear}.
Moreover, this result is recovered exactly by the adjoint method in a linear reciprocal system (as derived in Appendix~\ref{app:adjoint}):
\begin{equation}\label{eq:loss_weight_update}
    \frac{d J}{d \theta_i} =  -2\operatorname{Re}\{p^{(0)\top}|i\rangle \langle i |p^{(a)}\} = -2\operatorname{Re}\{p_i^{(0)}p_i^{(a)}\},
\end{equation}
where, with the frozen nudge and by linearity, $-\delta p/\beta$ is the response to the adjoint excitation alone, which can therefore be measured directly with $E^{(0)}$ switched off:
\begin{equation}
    p^{(a)} = -\frac{\delta p}{\beta} = M^{-1}\left.\frac{\partial C}{\partial p}\right|_{p^{(0)}}.
\end{equation}

The nudging term in Eq.~(\ref{eq:nudging}) is physically realizable provided the output dipoles (the support of $W$) can be driven by external sources with controlled amplitude and phase.
The adjoint is run on the same reciprocal system, driven by the phase-conjugated error, while the update requires phase-resolved measurements of $p_i$ at every tunable dipole, referenced to a common phase.
 
In summary, the training algorithm is:
\begin{itemize}
    \item \textbf{Free phase:} detune the transmitters with the usual inputs; measure $p^{(0)}$.
    \item \textbf{Nudged phase:} turn off the free-phase drive, $E^{(0)}=0$. The receivers are driven with the adjoint excitation $E^{(a)}=\left. \frac{\partial C}{\partial p}\right|_{p^{(0)}}$; measure $p^{(a)}$.
    \item Update each $\theta_i$ using Eq.~(\ref{eq:loss_weight_update}) and a chosen optimizer (SGD, Adam, etc.).
\end{itemize}
 
Two things are important to notice: the adjoint excitation $E^{(a)}=\partial C/\partial p|_{p^{(0)}}$ specifies what to apply at the receivers, while Eq.~(\ref{eq:loss_weight_update}) provides a simple, local gradient update for the physical neural network weights.
 
\section{Results}
\subsection{Physics simulation and neural network implementation} \label{sec:setup}
 
In this work, we focus on classification tasks using the well-known MNIST \cite{Lecun1998} and Fashion-MNIST \cite{Xiao2017} datasets.
We believe that there are no limitations on task complexity; it only depends on the total number of available parameters.
This work is a numerical proof of concept for scattering neural networks, and a real physical neural network would be highly desirable to improve on our current performance.
The results are simulations obtained with the code accessible at \url{https://github.com/TTimTT/ddep}.
 
To perform gradient descent on the weights, we use the \emph{Adam}  optimizer~\cite{Kingma2017a} with the MSE loss and the corresponding nudging function (derivation in Appendix~\ref{app:nudging_derivation}):
\begin{align}
    J(y, \hat{y}) =& ||y-\hat{y}||_2^2 \leftrightarrow C(\theta, y, p) = ||y-|Wp|^2||_2^2  \\
    \left. \frac{\partial C}{\partial p}\right|_{p^{(0)}} =& -2W^\top\left[\left(y - |W p^{(0)}|^2\right)\odot W p^{(0)*}\right]
\end{align}
 
\begin{figure}[H]
    \centering
    \includegraphics[width=0.4\textwidth]{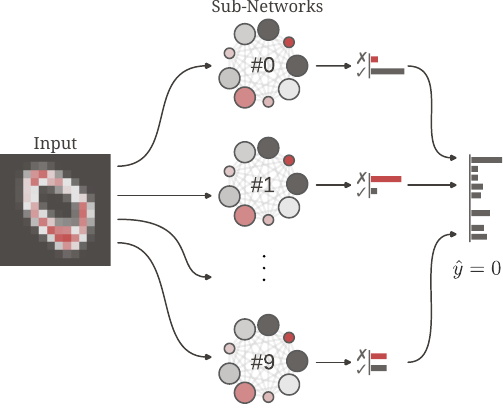}
    \caption{
        \small{
                 Winner-takes-all strategy: the same input is encoded in the tunability of 10 sub-networks, each with only two receivers. One receiver is mapped to a positive answer and the other to a negative one. The prediction is the index of the sub-network with the strongest positive answer.
            }
        }
    \label{fig:setup}
\end{figure}

Solving the physical propagation given by Eq.~(\ref{eq:cpldip-sol}) is the most power- and time-consuming task (representing about $97.8\%$ of the total execution time, see Appendix~\ref{app:impl_profiling}) and scales in complexity as $\mathcal{O}(N^3)$  \cite{laug}. 
Of course, this computation is performed passively in a real physical system, at a speed comparable to the speed of light.
Moreover, `the reference model has a total of 7850 trainable parameters for the MNIST'\cite{Wu2020}\textemdash corresponding to at least 7850 dipoles in our system\textemdash which is practical neither in our simulation nor experimentally.
Even though it is possible to have a single large network with 10 receivers, simulating this system is computationally prohibitive.
It would require approximately $7000$ scattering dipoles in total to match the MLP baseline's parameter count.
With the $\mathcal{O}(N^3)$ cost of the linear solve, this is roughly $(7000/843)^3/10\approx60$ times more computation, i.e., about 3 days per epoch instead of $4048$s.
The experimental implementation would also imply managing over 7000 signals at the same time, which is feasible but should ideally be done with integrated technologies.
To bring training down to about one hour per epoch, we reduced the input image size from $28\times 28$ to $14\times 14$ pixels.
We also implemented a \textit{winner-takes-all} \cite{Maass2000} strategy, as illustrated in Fig.~\ref{fig:setup}, where dedicated sub-networks make binary decisions, one per digit.
This drastically reduces the size of the matrices $M$ involved during training.
 
\begin{figure}[H]
    \centering   
    \includegraphics[width=0.4\textwidth]{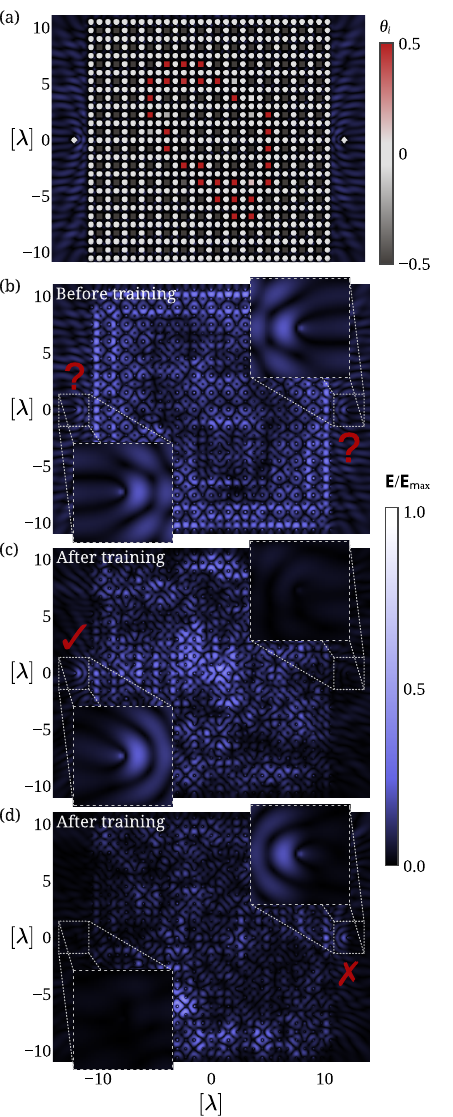}
    \caption{
        \small{
                \textbf{(a)} Grid layout of the dipole array, with the two receiving antennas (diamonds) on the outside, the input array (squares), and the trainable scatterers (disks). At initialization, the scatterer weights (detunings) are not yet trained, so their detuning is 0.
                \textbf{(b)} Normalized electric field amplitude map before training, for any sub-network.
                \textbf{(c)} Normalized electric field after training for sub-network 0. The field is stronger on the left receiver, indicating a positive answer.
                \textbf{(d)} Normalized electric field after training for sub-network 1. The field is stronger on the right receiver, indicating a negative answer.
            }
        }
    \refstepcounter{subfigure}\label{fig:setup_trained.a}
    \refstepcounter{subfigure}\label{fig:setup_trained.b}
    \refstepcounter{subfigure}\label{fig:setup_trained.c}
    \refstepcounter{subfigure}\label{fig:setup_trained.d}
    \label{fig:setup_trained}
\end{figure}

Accordingly, the same input is encoded in the 10 sub-networks, each specialized in single-digit recognition.
The final prediction corresponds to the index of the sub-network with the highest confidence\textemdash the highest measured intensity.
Each sub-network has two receivers, primarily for robustness against noise and to simplify the prediction.
 
In the single-receiver configuration, the wave intensity alone does not admit a universal threshold for binary classification across all sub-networks.
The solution is straightforward: by comparing the signal intensities of the two receivers, the system outputs a negative prediction when the right receiver exhibits greater intensity, and a positive prediction otherwise. This resembles the Forward-Forward (FF) learning algorithm \cite{Hinton2022}, though a key distinction emerges in terms of scalability. The FF algorithm requires a number of inference passes that grows linearly with the number of classes, whereas our framework circumvents this limitation entirely—requiring only a single inference pass regardless of class count.
 
Fig.~\ref{fig:setup_trained.a} depicts the array of dipoles arranged in a grid of $29\times 29$ dipoles spaced by $0.75\lambda$.
There are $N_t=196$ transmitters and $N_s=645$ scatterers, and the last $N_r = 2$ dipoles are the receivers, located on the sides of the grid array, totaling $N=843$ dipoles for each sub-network.
As we define 10 different sub-networks, the total number of parameters of our neural network is therefore $10\times(645+2) = 6470$.
In Fig.~\ref{fig:setup_trained.a}, squares represent the input dipoles, disks the trainable scatterers, and diamonds the two receivers.
The markers' color represents the detuning with respect to the working frequency.
Before training, the electric field intensity measured at the two receivers is almost identical, as shown in Fig.~\ref{fig:setup_trained.b}.
Figs.~\ref{fig:setup_trained.c} and ~\ref{fig:setup_trained.d} show the normalized electric field distribution obtained after 3 epochs of training for sub-network \#0 and sub-network \#1, respectively.
We see that, for the input image representing a 0, the first sub-network correctly redirects the electric field onto the left receiver, yielding a positive prediction.
In the second sub-network, the electric field is redirected onto the right receiver, which is interpreted as a negative prediction.
 
\subsection{Baseline results}
 
We start with a hyperparameter search on the learning rate and batch size using \emph{Optuna} \cite{Akiba2019} to establish an optimal baseline score (see Appendix~\ref{app:baseline_results} for detailed results).
We used a batch size of 32 and a learning rate of 0.034 for the rest of the work.
Figs.~\ref{fig:baseline.a} and ~\ref{fig:baseline.b} show the test accuracy and loss for both datasets over 3 epochs of training.
In the case of the MNIST dataset, we reached a plateau at $\sim 94\%$ test accuracy and a loss below 0.02.
This is comparable to the baseline performance of a simple multilayer perceptron (MLP) ($196\rightarrow 40 \rightarrow 10$), with similar input size and number of parameters, which also reaches $\sim 94\%$ accuracy.
These figures suggest that our system operates in a competitive performance regime for a proof-of-concept physical implementation.
For Fashion-MNIST, our results are close to those of a standard MLP, with $\sim 81\%$ test accuracy and a loss below $0.04$, and could be improved with a bigger network, i.e., more available weights.
The confusion matrices for both tasks are shown in Figs.~\ref{fig:baseline.c} and ~\ref{fig:baseline.d}, where class 6 of Fashion-MNIST corresponds to the shirt class, known to be the most difficult to classify correctly \cite{Ahmad2026}.
 
\begin{figure}[t]
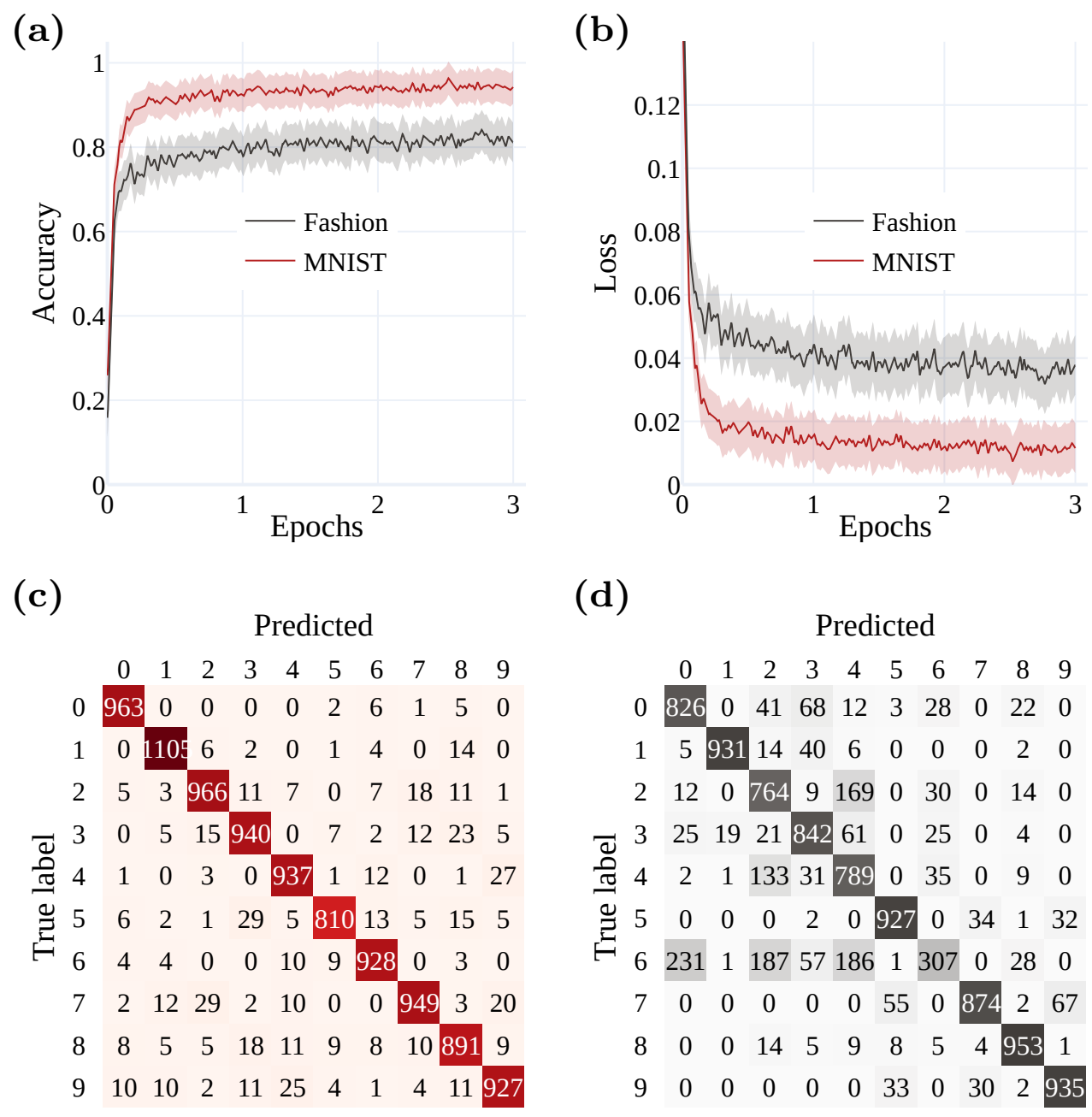

    \centering
    \customcaption{width=0.23\textwidth}{"./figures/fig_3a"}{(a)}{fig:baseline.a}{black}
    \customcaption{width=0.23\textwidth}{"./figures/fig_3b"}{(b)}{fig:baseline.b}{black}
 
    \customcaption{width=0.23\textwidth}{"./figures/fig_3c"}{(c)}{fig:baseline.c}{black}
    \customcaption{width=0.23\textwidth}{"./figures/fig_3d"}{(d)}{fig:baseline.d}{black}
    \caption{\small{
    \textbf{(a)} Test accuracy on the MNIST and Fashion-MNIST datasets during training for the baseline configuration.
    \textbf{(b)} Test loss on the MNIST and Fashion-MNIST datasets during training for the baseline configuration.
    \textbf{(c)} MNIST confusion matrix.
    \textbf{(d)} Fashion-MNIST confusion matrix.
    }}
    \label{fig:baseline}
\end{figure}
 
\subsection{Noise and disorder analysis}\label{sec:result-noise}
 
We also studied the effects of dipole-moment measurement noise and of the dipoles' spatial disorder, during training and inference, on the MNIST digits dataset.
We performed an Optuna grid sweep over different levels of uniform complex additive measurement noise:
 
\begin{equation}
    \tilde{p} = p + (\epsilon_{\Re} + j \epsilon_{\Im})
\end{equation}
 
and of uniform additive positional disorder of the dipoles:
\begin{equation}
    \tilde{r} = \vec{r} + \epsilon_{\vec{r}}
\end{equation}
 
The random variables $\epsilon_{\Re}, \epsilon_{\Im}$ are drawn from a uniform distribution on the interval $[-\sigma_p, \sigma_p]$.
Similarly, the random variable $\epsilon_{\vec{r}}$ is drawn from a uniform distribution on the interval $[-\sigma_{\vec{r}}, \sigma_{\vec{r}}]$:
 
$$\epsilon_{\Re}, \epsilon_{\Im} \sim \mathcal{U}(-\sigma_p, \sigma_p) \text{ and } \epsilon_{\vec{r}}\sim \mathcal{U}(-\sigma_{\vec{r}}, \sigma_{\vec{r}}).$$
 
We used $\sigma_p\in \{0, 0.0001, 0.001, 0.01, 0.1\}$ and $\sigma_{\vec{r}}\in \{0, 0.0001, 0.001, 0.01, 0.1, 0.3, 0.6\}$, yielding $35$ grid points in the search space.
The results for additive measurement noise are shown in Fig.~\ref{fig:noise-study.a},
where the color scale is the mean noise value for each grid point, $\langle \sigma_i \rangle$, normalized by the mean noise-free dipole moment $\langle |p| \rangle$:
 
\begin{equation}
   L_{\sigma_{p,i}} = \frac{ \sigma_{p,i} }{\langle |p| \rangle}
\end{equation}
 
Measurement noise degrades the test accuracy from $L_{\sigma_{p,i}}\approx 4\%$, where the final accuracy drops to about 75\%.
At 40\% noise, training fails and the accuracy stays close to chance level.
However, since our system is reciprocal and passive in nature, the noise can be compensated by increasing the signal power in the antennas, as shown theoretically in \cite{Bosch2026}.
 
In the case of spatial disorder, we normalized by the standard dipole spacing of $0.75\lambda$:
 
\begin{equation}
   L_{\sigma_{{\vec{r}},i}} = \frac{ \sigma_{{\vec{r}},i} }{0.75}
\end{equation}
 
Fig.~\ref{fig:noise-study.b} shows that, for all disorder levels tested, the test accuracy during training is largely unaffected.
Even with a disorder level of $L_{\sigma_{{\vec{r}}}} = 80\%$, as shown in Fig.~\ref{fig:noise-study.c}, the network was able to reach a test accuracy of $\sim 91\%$.
In addition, we present the correlation matrix in Fig.~\ref{fig:noise-study.d}, which shows a correlation of $-0.08$ between the final test accuracy and the level of spatial disorder.
This indicates that training remains effective in strongly disordered arrays.
Since the gradient is measured on the physical system itself, it is exact for whichever configuration is realized, and no knowledge of the actual positions, hence of $M$, is required: static manufacturing imperfections are absorbed by the training rather than degrading it.

\begin{figure}[t]
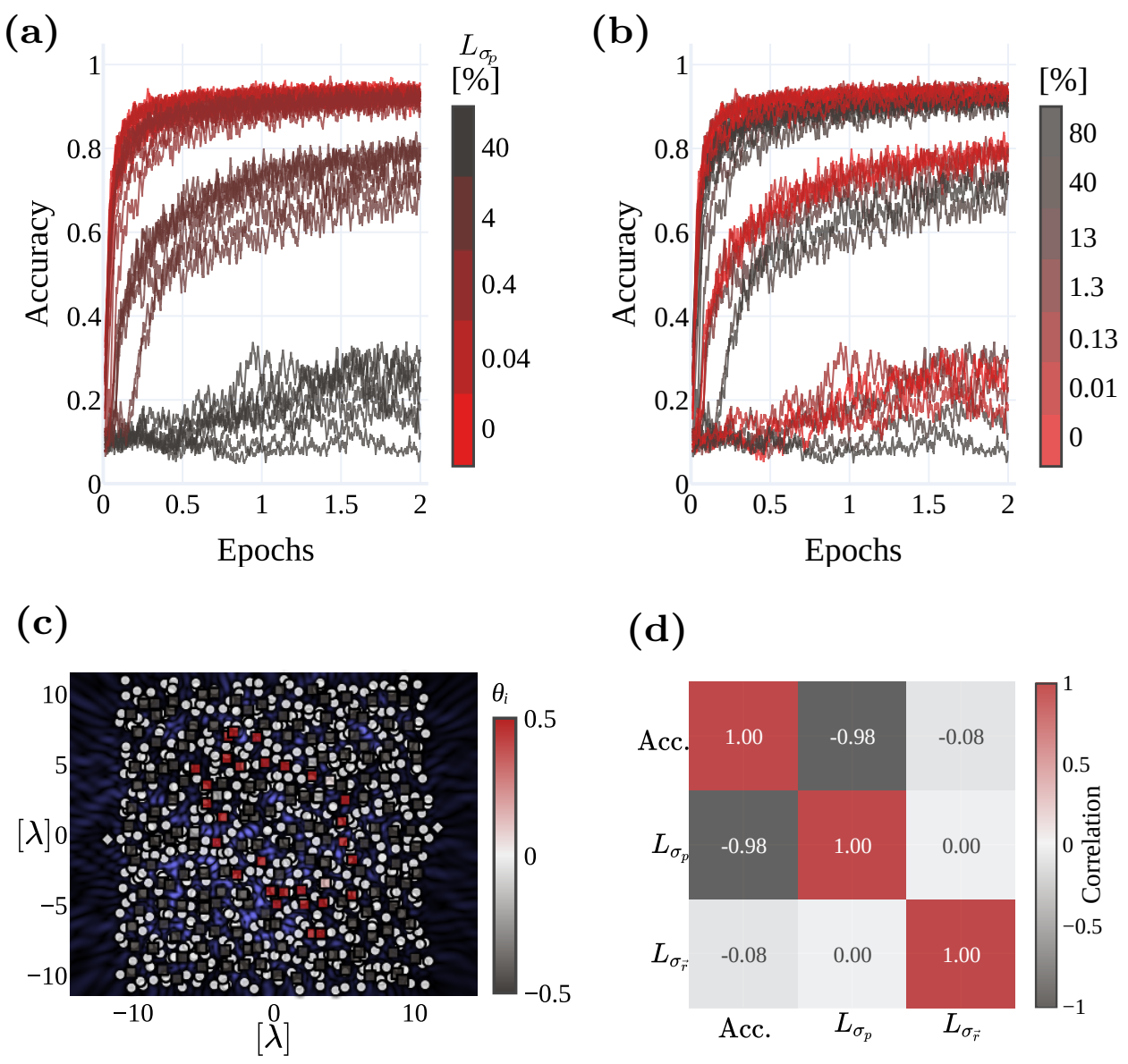

    \centering
    \customcaption{width=0.23\textwidth}{"./figures/fig_4a"}{(a)}{fig:noise-study.a}{black}
    \customcaption{width=0.23\textwidth}{"./figures/fig_4b"}{(b)}{fig:noise-study.b}{black}
 
    \customcaption{width=0.24\textwidth}{"./figures/fig_4dbis"}{(c)}{fig:noise-study.c}{black}
    \customcaption{width=0.21\textwidth}{"./figures/fig_4c"}{(d)}{fig:noise-study.d}{black}
\caption{
    \small{
        \textbf{(a)} Test accuracy for various levels of uniform additive measurement noise, with $L_{\sigma_{p,i}}$ between 0 and $40\%$. The higher the noise, the worse the final test accuracy.
        \textbf{(b)} Test accuracy for various levels of dipole positional disorder, with $L_{\sigma_{\vec{r},i}}$ between 0 and $80\%$. Each disorder level includes runs at all noise levels; the intermediate and low-accuracy curves correspond to $L_{\sigma_{p,i}}=4\%$ and $40\%$, respectively. At a given noise level, the test accuracy is largely unaffected by the disorder.
        \textbf{(c)} Example of sub-network \#0's dipole array with a spatial disorder of $L_{\sigma_{\vec{r},i}} = 80\%$.
        \textbf{(d)} Correlation matrix of the test accuracy, measurement noise, and disorder. There is a strong anticorrelation between noise level and test accuracy, and a correlation of $-0.08$ between dipole disorder and test accuracy. The final accuracy is not significantly influenced by spatial disorder.
    }
}
    \label{fig:noise-study}
\end{figure}
 
\section{Conclusion}
 
We have presented an adapted Equilibrium Propagation learning algorithm for networks of coupled electric dipoles in two-dimensional electromagnetism.
Although the physical steady state of such systems corresponds to a driven-dissipative, non-equilibrium regime, Eq.~(\ref{eq:cpldip-sol}) can be recovered as the stationary point of a  functional
$\mathcal{E}(\theta, p) = \frac{1}{2}p^\top M(\theta)p - p^\top E$, whose stationarity condition coincides with the coupled-dipole equations.
This construction connects the problem to a two-phase learning protocol based on EP.
We showed how to adapt this rule to intensity readout, for which the cost is a real-valued, non-holomorphic function of the complex state and existing holomorphic formulations of EP do not apply.
For a holomorphic energy and any real-valued cost, the nudged phase captures only the holomorphic part of the gradient, while the missing antiholomorphic part is its complex conjugate, so that the exact gradient is recovered by taking the real part of the EP estimator.
For linear reciprocal systems, this identity reduces to the \textit{in situ} adjoint gradient, and it extends formally to antennas whose nonlinear response is holomorphic in the dipole moment (Appendix~\ref{app:nonlinear}).
 
Furthermore, interpreting the real part of the inverse polarizability as a tunable antenna detuning provides a physically meaningful set of trainable parameters that naturally respect the passivity constraints imposed by the optical theorem.
These parameters enter the interaction operator only through its diagonal, such that each weight update reduces to the product of two currents measured on the corresponding antenna, in a free phase and in an adjoint phase, without knowledge of the scattering operator, even in the case of a antenna-local nonlinearity described by a holomorphic functional.
We exploited structural nonlinearity by encoding the input data directly in the tunability of the emitting antennas.
We showed numerically that such a system can be trained to achieve over $94\%$ test accuracy on the MNIST classification task, and to about 91\% under strong positional disorder (80\% of the lattice spacing)..
Because the gradient is measured on the system itself, static imperfections are absorbed by training, which makes the scheme well suited to imperfect physical implementations.
Several natural extensions of this work can be envisioned.
On the theoretical side, the framework could be generalized to three-dimensional and vectorial electromagnetic systems.
Another promising direction is the incorporation of time-dependent or nonlinear elements, in particular non-holomorphic, intensity-dependent nonlinearities such as Kerr-type responses, which fall outside the present formulation.
Although the training tasks considered here did not reveal any particular limitation in expressivity, such elements could shed light on the role of nonlinearity in training accuracy.
On the experimental side, tunable antenna arrays operating at radio frequencies offer a direct route to a physical implementation of the proposed scheme.
Finally, an optical implementation would be highly attractive for realizing fast, energy-efficient, and large-scale physical neural networks; achieving this, however, will require adapting the algorithm to cope with the challenge of phase-resolved measurements in optical systems.
 
\section*{ACKNOWLEDGMENTS}
This work was supported by the SNSF (grant number 212577) through the project `Ultra-Compact Non-linear Metamaterial Wave Processors for Analog Deep Learning'.
The authors thank Tristan Nerson, Henri Leroy, and Mohammad Tavakkoli for their valuable feedback on this work. R. Fleury acknowledges useful discussions with Benjamin Scellier, Philip del Hougne, and Cyrill Bösch on model-free training methods related to EP and adjoint methods.
 
\section*{DATA AVAILABILITY STATEMENT}
The source code for the simulations presented in this study is openly accessible under the terms of the GNU General Public License v3.0 (GPLv3) and can be retrieved from \url{https://github.com/TTimTT/ddep}.
 
The simulation results and underlying data supporting the findings of this study are available from the corresponding author upon request.
 
\newpage
\bibliographystyle{apsrev4-2}
\bibliography{apssamp}
 
\newpage
\appendix
 
\section{Equilibrium propagation with non-holomorphic readout}\label{app:EP_nonholo}
 
For any smooth function along a path parametrized by a real variable, the Wirtinger chain rule gives:
\begin{equation}\frac{\mathrm{d}J}{\mathrm{d}\theta}=\left(\frac{\partial C}{\partial p}\right)^\top\frac{\partial p^{(0)}}{\partial \theta}+\left(\frac{\partial C}{\partial p^*}\right)^\top\frac{\partial p^{(0)*}}{\partial \theta}.
\end{equation}
Two facts make the second term the conjugate of the first.
First, $C$ is real, so:
\begin{equation}
    \frac{\partial C}{\partial p^*} = \left(\frac{\partial C}{\partial p}\right)^*,
\end{equation}
Second, $\beta$ and $\theta$ are real, so:
\begin{equation}
\frac{\partial p^*}{\partial \beta} = \left(\frac{\partial p}{\partial \beta}\right)^*, \frac{\partial p^*}{\partial \theta} = \left(\frac{\partial p}{\partial \theta}\right)^*,
\end{equation}
and we define $Q = \left(\frac{\partial C}{\partial p}\right)^\top \frac{\partial p^{(0)}}{\partial \theta}$ such that:
\begin{equation}\label{eq:loss_Q}
    \frac{\mathrm{d}J}{\mathrm{d}\theta}=Q+Q^* 
\end{equation}
 
We now differentiate $F(\theta, y, \beta, p)$ with respect to $\beta$ and $\theta$ using Wirtinger derivatives.
Let $\Phi(\theta,\beta)\equiv F\big(\theta,y,\beta,p^{(\beta)}(\theta)\big)$ be $F$ evaluated along the nudged state $p^{(\beta)}$, and $f(\theta,\beta)\equiv\frac{\partial F}{\partial\theta}\big(\theta,y,\beta,p^{(\beta)}(\theta)\big)$ its explicit $\theta$-derivative along the same path:
\begin{align}
\frac{\partial \Phi}{\partial \beta} &= C + \left(\frac{\partial F}{\partial p}\right)^{\!\top}\frac{\partial p}{\partial \beta} + \left(\frac{\partial F}{\partial p^*}\right)^{\!\top}\frac{\partial p^*}{\partial \beta} \\
\frac{\partial \Phi}{\partial \theta} &= \frac{\partial F}{\partial \theta} + \left(\frac{\partial F}{\partial p}\right)^{\!\top}\frac{\partial p}{\partial \theta} + \left(\frac{\partial F}{\partial p^*}\right)^{\!\top}\frac{\partial p^*}{\partial \theta},
\end{align}

Recall that $\mathcal{E}$ is holomorphic in $p$, whereas $C$ depends on both $p$ and $p^*$.
Along the nudged state, $\partial F/\partial p=0$ and $\partial F/\partial p^*=\beta\,\partial C/\partial p^*$, so:
\begin{align}
\frac{\partial \Phi}{\partial \beta} &= C + \beta\left(\frac{\partial C}{\partial p^*}\right)^\top\frac{\partial p^*}{\partial \beta} \\
\frac{\partial \Phi}{\partial \theta} &= f + \beta\left(\frac{\partial C}{\partial p^*}\right)^\top\frac{\partial p^*}{\partial \theta},
\end{align}
 
Since $p^{(\beta)}(\theta)$ is a smooth function of $(\theta,\beta)$, so is $\Phi$, and its mixed derivatives commute:
\begin{equation}
    \partial_{\theta}\partial_\beta \Phi=\partial_\beta\partial_{\theta} \Phi
\end{equation}
Evaluating the last expression at $\beta=0$ therefore gives:
\begin{equation}
    \left.\dfrac{d C }{d\theta}\right|_{\beta=0} = \partial_\beta f + \left(\frac{\partial C}{\partial p^*}\right)^\top\frac{\partial p^{(0)*}}{\partial \theta}.
\end{equation}
From Eq.~(\ref{eq:loss_cost}) and Eq.~(\ref{eq:loss_Q}), we obtain the gradient of the loss function:
\begin{equation}\label{eq:loss_FQ}
\dfrac{d J}{d\theta} = \partial_\beta f + Q^*.
\end{equation}
Subtracting Eq.~(\ref{eq:loss_FQ}) from Eq.~(\ref{eq:loss_Q}) yields:
\begin{equation}
Q = \partial_\beta f
\end{equation}
Hence, we can write:
\begin{align}
\dfrac{d J}{d\theta} &= \partial_\beta f + \left(\partial_\beta f\right)^* \\
 &= 2\operatorname{Re}\left\{ \partial_\beta f\right\}.
\end{align}
We conclude that:
\begin{equation} \label{eq:EP_nonholo}
\dfrac{d J}{d\theta} = \lim_{\beta \to 0}\frac{2}{\beta}\operatorname{Re}\left\{\frac{\partial F}{\partial \theta}(\theta,y,\beta,p^{(\beta)}) - \frac{\partial F}{\partial \theta}(\theta,y,0,p^{(0)})\right\}.
\end{equation}
 
\section{Equilibrium propagation simplification}\label{app:EP_derivation}
Recall that the EP gradient is given by the formula:
\begin{equation}
    \frac{d J}{d \theta} = \lim_{\beta \to 0}\frac{2}{\beta}\operatorname{Re}\left\{\frac{\partial F}{\partial \theta}(\theta,y,\beta,p^{(\beta)}) - \frac{\partial F}{\partial \theta}(\theta,y,0,p^{(0)})\right\},
\end{equation}
 
or equivalently:
\begin{align}
    \frac{d J}{d \theta} &= \underbrace{\frac{\partial C}{\partial \theta}\bigg|_{p \text{ fixed}}}_{= 0} + \lim_{\beta\to 0}\frac{2}{\beta}\operatorname{Re}\left\{\frac{\partial \mathcal{E}}{\partial \theta}\bigg|_{p^{(\beta)}} - \frac{\partial \mathcal{E}}{\partial \theta}\bigg|_{p^{(0)}}\right\}  \\
    &= \lim_{\beta\to 0}\frac{2}{\beta}\operatorname{Re}\left\{\frac{\partial \mathcal{E}}{\partial \theta}\bigg|_{p^{(\beta)}} - \frac{\partial \mathcal{E}}{\partial \theta}\bigg|_{p^{(0)}}\right\} \\
    &=  \lim_{\beta \to 0} \frac{1}{\beta}\operatorname{Re}\left\{p^{(\beta)\top} \frac{\partial M}{\partial \theta}p^{(\beta)}
 - p^{(0)\top} \frac{\partial M}{\partial \theta}p^{(0)}\right\}
\end{align}

The EP formula computes the variation of $\mathcal{E}$ along $\delta p$. We write $p^{(\beta)} = p^{(0)} + \delta p$ to simplify the term:
\begin{align}
    &\qquad p^{(\beta)\top} \frac{\partial M}{\partial \theta}p^{(\beta)}
    - p^{(0)\top} \frac{\partial M}{\partial \theta}p^{(0)} \nonumber \\
    &= \left(p^{(0)} + \delta p\right)^\top \frac{\partial M}{\partial \theta} \left(p^{(0)} + \delta p\right) - p^{(0)\top} \frac{\partial M}{\partial \theta} p^{(0)} \\
   &=\underbrace{p^{(0)\top} \frac{\partial M}{\partial \theta} p^{(0)}}_{\text{cancels}} + p^{(0)\top} \frac{\partial M}{\partial \theta} \delta p  \nonumber \\
    &\quad- \underbrace{p^{(0)\top} \frac{\partial M}{\partial \theta} p^{(0)}}_{\text{cancels}} +
    \delta p^\top \frac{\partial M}{\partial \theta} p^{(0)} + \underbrace{\delta p^\top \frac{\partial M}{\partial \theta} \delta p}_{O(\delta p^2)}
\end{align}
To first order in $\delta p$:
\begin{align}
    \frac{d J}{d \theta} &= \lim_{\beta \to 0} \frac{1}{\beta}\operatorname{Re}\left\{p^{(0)\top} \frac{\partial M}{\partial \theta} \delta p + \delta p^\top \frac{\partial M}{\partial \theta} p^{(0)}\right\}
\end{align}
 
Recall from Eq.~(\ref{eq:M_theta}) that $\frac{\partial M}{\partial \theta}$ is symmetric:
\begin{align}
     \frac{d J}{d \theta} &=\lim_{\beta\to 0}\frac{1}{\beta}2\operatorname{Re}\left\{ \delta p^{\top} \frac{\partial M}{\partial \theta}p^{(0)}\right\} \\
     &= -\lim_{\beta\to 0}\frac{1}{\beta}2\operatorname{Re}\left\{ \left(\beta M^{-1}\left. \frac{\partial C}{\partial p}\right|_{p^{(\beta)}}\right)^\top\frac{\partial M}{\partial \theta}p^{(0)} \right\} \\
     &= -2\operatorname{Re}\left\{\left. \frac{\partial C}{\partial p}\right|_{p^{(0)}}^\top M^{-1}\frac{\partial M}{\partial \theta}p^{(0)} \right\} 
\end{align}
 
where, in the last step, we used the continuity of the nudging in $\beta$.
Finally,  Eq.~(\ref{eq:M_theta}) simplifies the relation to:
\begin{equation}
     \frac{d J}{d \theta_i} = -2\operatorname{Re}\left\{\left. \frac{\partial C}{\partial p}\right|_{p^{(0)}}^\top M^{-1}|i\rangle\langle i|p^{(0)} \right\} 
\end{equation}
 
In the end, $\beta$ drops out.
 
\section{Adjoint gradient update derivation \label{app:adjoint}}
The gradient of the loss can be expressed using Wirtinger derivatives under the following conditions:
\begin{itemize}
    \item The loss $J$ is a real scalar function: $J: \mathbb{R}^{N_{r}}\times \mathbb{R}^{N_r}\longrightarrow \mathbb{R}$.
    \item The cost $C$ is a multivariate function:
    
    $C:  \mathbb{R}^{N}\times \mathbb{R}^{N_r} \times\mathbb{C}^{N}\longrightarrow \mathbb{R}$.
    \item The dipole moment $p$ is a complex vector that depends implicitly on the real parameters $\theta$.
\end{itemize}
 
In element-wise notation, the Wirtinger chain rule then gives
\begin{align}
    \frac{d J}{d \theta_i} &= \left.\dfrac{d C}{d \theta_i}\right|_{p^{(0)}}\\
    &=  \left.\sum_{k=1}^N \frac{\partial C}{\partial p_k}\frac{\partial p_k}{\partial \theta_i} + \frac{\partial C}{\partial p^*_k}\frac{\partial p^*_k}{\partial \theta_i}\right|_{p^{(0)}} \\
    &= 2\operatorname{Re}\left\{\left. \sum_{k=1}^N\frac{\partial C}{\partial p_k} \frac{\partial p_k}{\partial \theta_i} \right|_{p^{(0)}} \right\} \\
    &= 2\operatorname{Re}\left\{\left.\frac{\partial C}{\partial p}^\top \frac{\partial p}{\partial \theta_i}\right|_{p^{(0)}}\right\}. \label{eq:der_loss}
\end{align}
 
Moreover, the drive $E^{(0)}$ is set in such a way that it does not depend on the parameters $\theta$.
Its only role is to inject nonzero electromagnetic energy into the system.
We obtain:
 
\begin{align}
    \frac{\partial E^{(0)}}{\partial \theta_i} &= 0  \Rightarrow \frac{\partial M}{\partial\theta_i} p^{(0)} + \left.M\frac{\partial p}{\partial\theta_i}\right|_{p^{(0)}} = 0, \\
    \left.\frac{\partial p}{\partial\theta_i}\right|_{p^{(0)}}&=-M^{-1}\frac{\partial M}{\partial \theta_i}p^{(0)},\ \text{with } \frac{\partial M}{\partial \theta_i}=|i\rangle\langle i| \\
    &=-M^{-1}|i\rangle\underbrace{\langle i|p^{(0)}}_{p^{(0)}_i} = -M^{-1}|i\rangle p^{(0)}_i \label{eq:partial_pi}
\end{align}
 
Using Eq.~(\ref{eq:partial_pi}) in Eq.~(\ref{eq:der_loss}) yields:
 
\begin{equation}\label{eq:loss_find_nudge}
    \frac{d J}{d\theta_i} = 2\operatorname{Re}\left\{- \left.\frac{\partial C}{\partial p}\right|_{p^{(0)}}^\top M^{-1}|i\rangle p^{(0)}_i\right\}.
\end{equation}
 
From this last equation, we observe that minimizing the loss $J(y, \hat{y})$ without knowledge of our operator $M$ is possible if we define an adjoint excitation $E^{(a)}$ equal to $\left.\frac{\partial C}{\partial p}\right|_{p^{(0)}}^\top$:
\begin{equation}
    \left.\frac{\partial C}{\partial p}\right|_{p^{(0)}}^\top = E^{(a)\top} = p^{(a)^\top}M^\top = p^{(a)^\top}M,
\end{equation}
 
where the last equality follows from the symmetry of $M$, established in Sec.~\ref{sec:cpdl}.
Substituting into Eq.~(\ref{eq:loss_find_nudge}) yields
\begin{align}
    \frac{d J}{d \theta_i} &= -2\operatorname{Re}\left\{ p^{(a)^\top}M M^{-1}|i\rangle p^{(0)}_i\right\} \\
    &= -2\operatorname{Re}\left\{ p_i^{(a)} p^{(0)}_i\right\}.
\end{align}
 
\section{Formal extension to holomorphic nonlinear antennas}\label{app:nonlinear}
In this appendix, we show that the training scheme of Sec.~\ref{sec:EP} extends formally to nonlinear antennas, provided their response is a holomorphic function of the dipole moment.
We discuss the physical scope of this condition at the end of the appendix.
We replace the linear constitutive relation of each antenna by
\begin{equation}
    \alpha_i^{-1}p_i \longrightarrow \alpha_i^{-1}p_i + \varphi_i(p_i),
\end{equation}
where $\varphi_i$ is holomorphic and does not depend on the trainable parameters $\theta$.
Let $V_i$ be a primitive of $\varphi_i$, i.e., $\frac{dV_i}{d p_i} = \varphi_i$.
The energy functional of Eq.~(\ref{eq:potential}) becomes
\begin{equation}\label{eq:potential_nl}
    \mathcal{E}(\theta, p) = \frac{1}{2}p^\top M p + \sum_{i=1}^{N}V_i(p_i) - p^\top E,
\end{equation}
which is still holomorphic in $p$.
Its stationarity condition gives the nonlinear state equation
\begin{equation}\label{eq:state_nl}
    \frac{\partial \mathcal{E}}{\partial p} = 0\Rightarrow Mp + \varphi(p) = E,
\end{equation}
where $\varphi(p) = (\varphi_1(p_1),\ldots,\varphi_N(p_N))^\top$.
 
Unlike in the linear case, Eq.~(\ref{eq:state_nl}) cannot be solved by a matrix inversion and may admit several solutions.
As is standard in EP, we assume that the free state $p^{(0)}$ is a stable, isolated solution.
 
The Jacobian of the vector field $Mp + \varphi(p) - E$ is
\begin{equation}\label{eq:K_nl}
    K(p) \equiv M + \operatorname{diag}\big(\varphi'(p)\big),
    \qquad \varphi'_i = \frac{\mathrm{d}\varphi_i}{\mathrm{d}p_i}.
\end{equation}
Since $M$ is symmetric and the nonlinear term is diagonal, $K$ is symmetric as well, consistent with the existence of the holomorphic functional~(\ref{eq:potential_nl}).
Moreover, $\varphi$ does not depend on $\theta$, so that $\partial K/\partial\theta_i = \partial M/\partial\theta_i = |i\rangle\langle i|$.
 
The derivation of Appendix~\ref{app:EP_nonholo} only requires $\mathcal{E}$ to be holomorphic and $C$ to be real, so Eq.~(\ref{eq:EP_nonholo}) still holds.
The nudged state $p^{(\beta)}$ now satisfies
\begin{equation}
    Mp^{(\beta)} + \varphi\left(p^{(\beta)}\right) - E^{(0)} + \beta\left.\frac{\partial C}{\partial p}\right|_{p^{(\beta)}} = 0,
\end{equation}
and, to first order in $\beta$,
\begin{equation}\label{eq:dp_nl}
    \delta p = p^{(\beta)} - p^{(0)} = -\beta\, K^{-1}\left.\frac{\partial C}{\partial p}\right|_{p^{(0)}} + \mathcal{O}(\beta^2),
\end{equation}
with $K$ evaluated at $p^{(0)}$.
 
Since $\frac{\partial \mathcal{E}}{\partial \theta_i} = \frac{1}{2}p_i^2$, Eq.~(\ref{eq:EP_nonholo}) gives
\begin{equation}\label{eq:grad_nl}
    \frac{dJ}{d\theta_i}
    = \lim_{\beta\to 0}\frac{1}{\beta}\operatorname{Re}\left\{\left(p_i^{(\beta)}\right)^2 - \left(p_i^{(0)}\right)^2\right\}
    = -2\operatorname{Re}\left\{p_i^{(0)}p_i^{(a)}\right\},
\end{equation}
where
\begin{equation}
    p^{(a)} = -\lim_{\beta\to0}\frac{\delta p}{\beta} = K^{-1}\left.\frac{\partial C}{\partial p}\right|_{p^{(0)}}.
\end{equation}
The same expression follows from the adjoint method of Appendix~\ref{app:adjoint}, with $M$ replaced by $K(p^{(0)})$.
 
The update rule therefore keeps the same local form as in the linear case, but the way it is measured differs in two respects.
First, the nudged phase must be performed with the input drive $E^{(0)}$ switched on, because $K$ depends on the operating point $p^{(0)}$; switching the transmitters off, as in the linear algorithm of Sec.~\ref{sec:EP}, would probe the response around a different state.
Second, $\beta$ must be small, since Eq.~(\ref{eq:dp_nl}) holds only to first order.
 
In practice, the finite-$\beta$ estimate of Eq.~(\ref{eq:grad_nl}) has an $\mathcal{O}(\beta)$ bias, which can be reduced to $\mathcal{O}(\beta^2)$ with a symmetric nudge:
\begin{equation}
    \frac{1}{2\beta}\operatorname{Re}\left\{\left(p_i^{(\beta)}\right)^2 - \left(p_i^{(-\beta)}\right)^2\right\},
\end{equation}
as in centered EP~\cite{Laborieux2021}.
 
For $\varphi = 0$, $K = M$, Eq.~(\ref{eq:dp_nl}) is exact for any $\beta$, and the linear algorithm of Sec.~\ref{sec:EP} is recovered.
 
Finally, we stress that this extension is formal, and that its physical scope is narrow.
For an antenna whose physics is invariant under time translation, delaying the drive by $\Delta t$ delays the response by the same amount, so that at a fixed frequency the steady-state response must be phase-covariant: $\varphi_i(e^{j\phi}p_i) = e^{j\phi}\varphi_i(p_i)$ for all real $\phi$.
Writing a holomorphic response as $\varphi_i(p_i)=\sum_n a_n p_i^n$, this condition imposes $a_n = 0$ for all $n\neq 1$: a holomorphic, phase-covariant response is necessarily linear.
The amplitude-dependent responses of time-invariant antennas, such as Kerr-type terms $\propto |p_i|^2 p_i$ or saturation, therefore always depend on $p_i^*$.
For these, $\mathcal{E}$ is no longer holomorphic, and the derivation of Appendix~\ref{app:EP_nonholo} does not apply.
Holomorphic nonlinear responses can nevertheless be obtained with actively phase-referenced elements, for instance circuits that mix the antenna signal with a local oscillator coherent with the drive.
Extending the present scheme to passive, $p^*$-dependent nonlinearities remains an open problem.
 
\section{Nudging derivation \label{app:nudging_derivation}}
 
There are two ways to obtain the nudging: either from its analytical form or by using the automatic differentiation capabilities of \emph{PyTorch}.
The prediction is obtained by measuring the intensity at the receivers in the free phase.
Let $g(p) = |Wp|^2$, such that $\hat{y} = g(p^{(0)})$.
In this work, we use the MSE loss:
 
\begin{equation}
    J(y, \hat{y}) = ||y-\hat{y}||_2^2 \leftrightarrow C(\theta, y, p) = ||y-|Wp|^2||^2_2
\end{equation}
 
Therefore, the nudging is given by:
\begin{equation}
    \left.\frac{\partial C}{\partial p}\right|_{p^{(0)}} = \left.\frac{\partial C}{\partial g}  \frac{\partial g}{\partial p}\right|_{p^{(0)}}
\end{equation}
 
\begin{align}
    \frac{\partial C}{\partial g_i} &= -2 (y_i-g_i),\qquad
    \frac{\partial g_i}{\partial p_k} = W_{ik} [W p^*]_i, \\
    \frac{\partial C}{\partial p_k} &= \sum_i \frac{\partial C}{\partial g_i}\frac{\partial g_i}{\partial p_k} \\
    &= - 2 \sum_i W_{ik} (y_i-g_i)  [W p^*]_i \\
    \left.\frac{\partial C}{\partial p}\right|_{p^{(0)}} &= -2W^\top\left[\left(y - |W p^{(0)}|^2\right)\odot W p^{(0)*}\right].
\end{align}
 
We can also use the cross-entropy loss function after normalizing $\hat{y}$ with the softmax function:
\begin{equation}
    \sigma(\hat{y})_k = \frac{e^{\hat{y}_k}}{\sum_{n=1}^{N_r} e^{\hat{y}_n}}.
\end{equation}
The loss and cost functions become:
\begin{align}
    J(y, \hat{y}) &= -\sum_{n=1}^{N_r} y_n\log(\sigma(\hat{y})_n) \\
    C(\theta, y, p) &= -\sum_{n=1}^{N_r} y_n\log(\sigma(g(p))_n)
\end{align}
Hence, we obtain the following nudging function:
\begin{align}
    &\frac{\partial C}{\partial g_i} = \sigma(g)_i - y_i\\
    &\frac{\partial C}{\partial p_k} = \sum_i \frac{\partial C}{\partial g_i}\frac{\partial g_i}{\partial p_k} =  \sum_i W_{ik} \left( \sigma(g)_i - y_i \right)  [W p^*]_i \\
    &\left.\frac{\partial C}{\partial p}\right|_{p^{(0)}} = W^\top \left[\left( \sigma(|W p^{(0)}|^2) - y \right)\odot Wp^{(0)*}\right]
\end{align}
 
In our code implementation, we use \emph{PyTorch} to compute this nudging.
We detach the $p^{(0)}$ tensor from the computation graph to avoid backpropagating through the full chain. Then, we apply the backward function to the loss.
This solution is more convenient and faster, and it yields a nudging function that is already evaluated at $p^{(0)}$.
 
\section{Code performance and profiling \label{app:impl_profiling}}
 
\begin{figure}[ht!]
    \centering
    \includegraphics[width=0.45\textwidth]{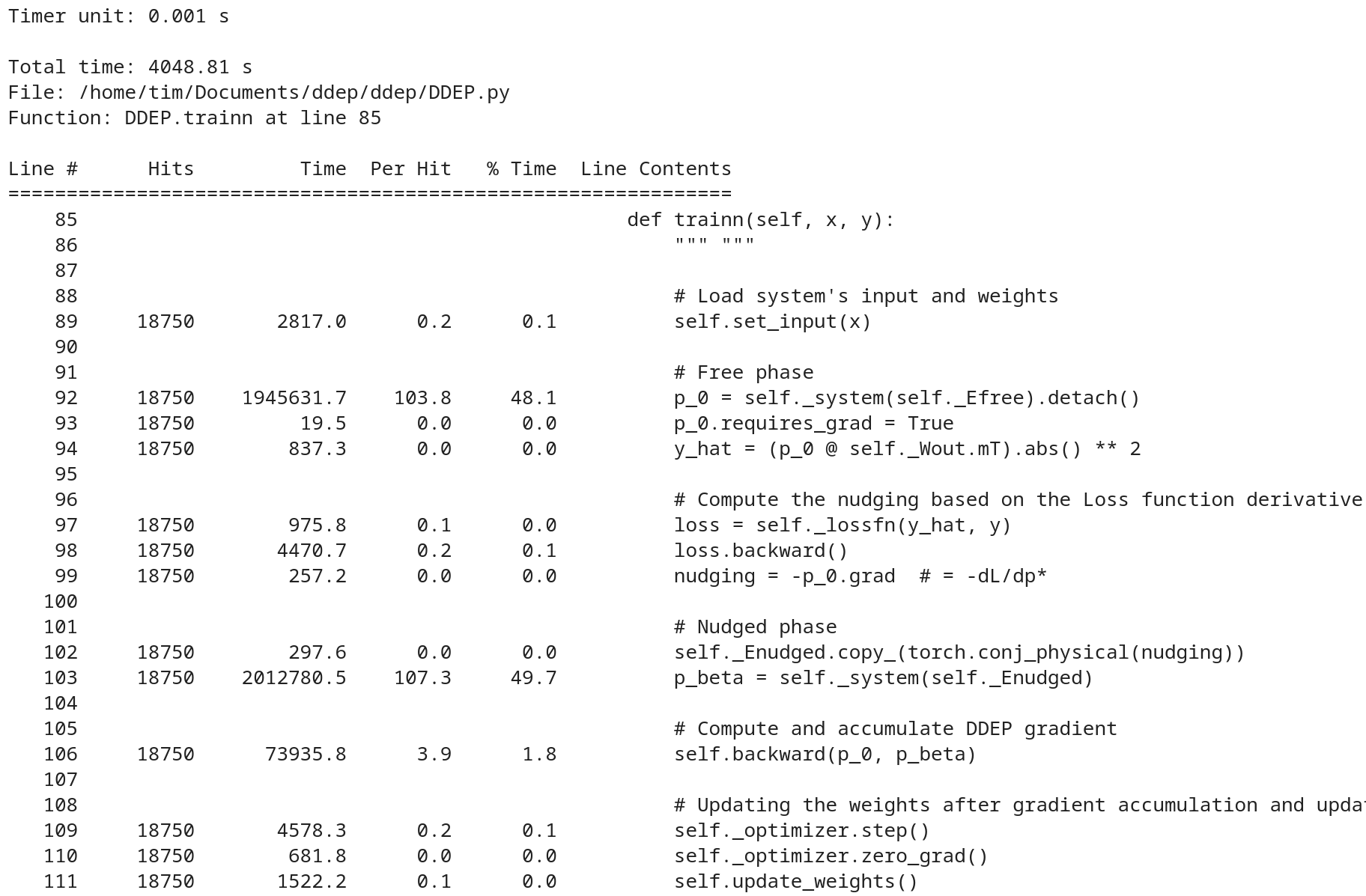}
    \caption{
        \small{
                Profiling results for a partial training of our model over one epoch, with a total execution time of $4048$s. Solving Eq.~(\ref{eq:compact-cpldip}) took 48.1\% of the computation time in the free phase and 49.7\% in the nudged phase, amounting to 97.8\% of the total time. Implementing such a network physically could reduce the partial training time to only $90$s. This profiling was done using a batch size of 32 and an input image size of [14, 14].
            }
        }
    \label{fig:profiling}
\end{figure}

The implementation relies heavily on \emph{PyTorch} libraries and is executed on an \emph{RTX 4080} GPU using CUDA. Eq.~(\ref{eq:compact-cpldip}) is solved as follows:
\begin{lstlisting}
    p = torch.linalg.solve(M, E)
\end{lstlisting}
 
Although this call uses \emph{cuSOLVER’s getrf}, the need to solve the linear system for each new input makes the physical simulation the bottleneck.
This is visible in Fig.~\ref{fig:profiling}, where $97.8\%$ of the total execution time is spent computing the dipole moments $p^{(0)}$ and $p^{(a)}$.
Implementing this neural network physically would reduce the training time of a single epoch from $4048$s to only $90$s.
 
\section{Hyperparameter search}\label{app:baseline_results}
 
\begin{figure}[H]
    \centering
    \includegraphics[width=0.45\textwidth]{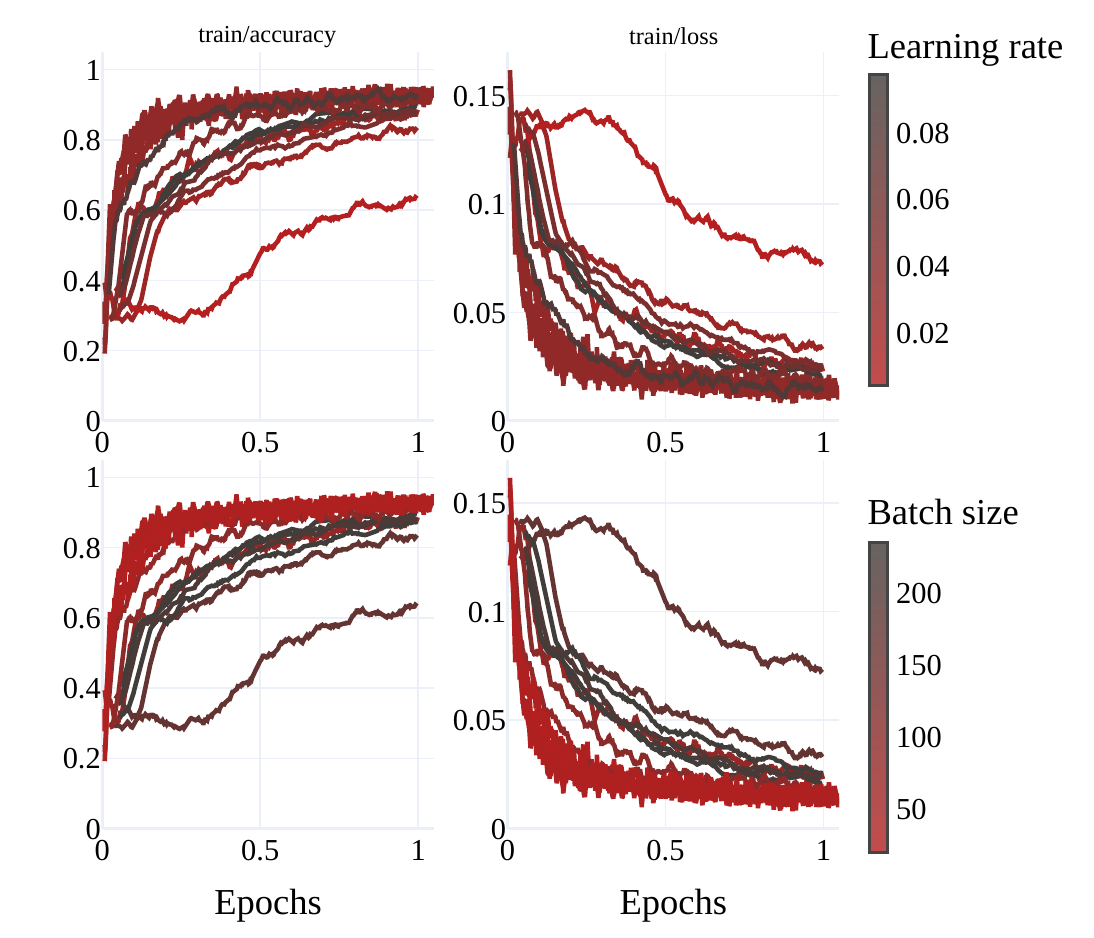}
    \caption{\small{Training accuracy and loss for different values of learning rate and batch size, used to identify the best set of hyperparameters. The search is performed with \emph{Optuna}'s TPESampler and pruning strategy. Each trial was run for a single epoch of training to avoid wasting time on unpromising hyperparameter values.
        }
    }
    \label{fig:app_baseline}
\end{figure}

\begin{figure}[H]
    \centering
    \includegraphics[width=0.4\textwidth]{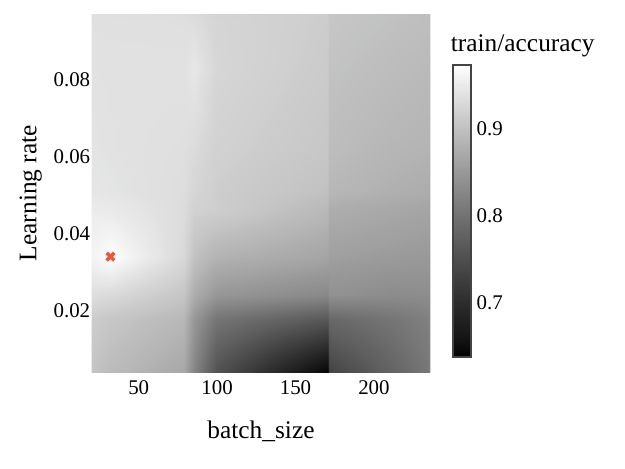}
    \caption{\small{2D heatmap of the training accuracy versus the learning rate and batch size. The whiter the color, the higher the training accuracy. From this analysis, we deduced our best set of hyper-parameters, marked with a cross: learning rate $\eta\approx0.034$ and batch size $=32$, which yield over $94\%$ training accuracy on MNIST.
        }
    }
    \label{fig:app_baseline_mat}
\end{figure}

In this section, we present the hyperparameter search performed with \emph{Optuna}.
More explicitly, we looked for the batch size and learning rate $\eta$ that maximize the training accuracy, as shown in Figs.~\ref{fig:app_baseline} and ~\ref{fig:app_baseline_mat}.
We concluded that a batch size of 32 and a learning rate $\eta\approx 0.034$ give the best training accuracy.
 
\end{document}